\documentstyle[12pt,amstex,amssymb] {article}
\begin {document}
\parindent=15pt
\begin{flushright}
{\bf US-FT/23-97}
\end{flushright}
\vskip .8 truecm
\begin{center}
{\bf NUCLEAR EFFECTS IN CHARMONIUM PRODUCTION IN QCD}\\ 

\vspace{.5cm}
C. Pajares, C.A. Salgado \\
\vspace{.5cm}
Departamento de F\'\i sica de Part\'\i culas, Universidade de Santiago de
Compostela, \\
E-15706-Santiago de Compostela, Spain \\
\vspace{.5cm}
and

\vspace{.5cm}
Yu.M. Shabelski \\
\vspace{.5cm}
Petersburg Nuclear Physics Institute, \\
Gatchina, St.Petersburg 188350 Russia \\
\end{center}
\vspace{1cm}

\begin{abstract}

\vskip 1. truecm
It is shown that the nuclear shadowing of charmonium due to the modification of
the nuclear parton distribution is similar in the factorization approach based
on non relativistic QCD and in the color evaporation model. In the first model,
a separate study of the color octet and color
singlet contributions to the yields of the various charmonium states as well as
the contributions of these states to the total $J/\Psi$ production is
performed.
It is found a clear $x_F$ dependence of these contributions which can reproduce
experimental data for moderate $x_F$. 

\end{abstract}


\vskip 1cm

July 1997.
\vskip 0.2truecm

{\bf US-FT/23-97}

\pagebreak

\section{INTRODUCTION}

Since the observation of a possible anomalous suppression 
in lead on lead
collisions by the NA50 collaboration \cite{NA501,NA502} 
a lot of theoretical work has
been done trying to explain the data. Some papers use conventional physics
\cite{Nestor}-\cite{Alf} but most of the explanations are based on QGP
formation \cite{DD}-\cite{qa}, the new state
that has been predicted to be formed at high enough temperatures and
densities of hadronic matter and from which $J/\Psi$ suppression has been proposed as a
signature long ago \cite{MS}.
Although
there are some consensus in the possibility of this state to be reached in 
$158$ AGeV/c lead on lead collisions, a lot of
uncertainties remain to be solved on the production and suppression of the 
$J/\Psi$. 
A deep study of the absorption mechanisms seems necessary in order to 
assert the anomaly on the observed suppression. 

Charmonium production in hadron-hadron collisions is now quite accurately
described by a factorization approach (FA)\footnote{Sometimes it is called Color
Octet Model, although it seems now more frequent FA. The name COM is
not very correct because it is not a model, but the solution of an effective
theory and not only color octet states are taken into account.} 
based on non-relativistic QCD
(NRQCD) \cite{BBL},
although the so called color evaporation model (CEM) 
has also a remarkable success in 
describing integrated as well as differential cross sections \cite{CEV}
\cite{halzen}.
The main difference between them is the treatment of color, being the CEM quite
radical in this sense: it simply forgets about color.
In the calculations of this paper both models will be used, turning out to give
almost identical results.

In parton model, the production mechanism is described by the fusion of a parton
coming from the projectile with longitudinal momentum fraction $x_1$ and
a parton from the target with longitudinal momentum fraction $x_2$. For a fixed
$x_F>0$ ($x_F\equiv x_1-x_2$), this momentum fraction are $x_1\simeq x_F$ and
$x_2\simeq M^2/x_1s$.

For 
collisions involving nuclei the situation is much more complex. 
Multiple scattering of the initial partons as well as of the 
produced pre-resonant states
with nucleons from the nucleus and even of the produced resonances with other
produced particles (co-movers) must be taken into account. It is very common to
call initial state interactions to multiple interactions of the 
initial partons
(mostly gluons in the case of $J/\Psi$ and $\Psi'$) and final state
interactions to the rescattering of the pre-resonant state. These names are
senseless at high energies, where time ordering is lost.
It is possible, however, to factorize the multiple scattering of 
light partons from that of the heavy pre-resonance. The former can be
described by modifying the nucleon structure function inside nuclei (EMC
effect). For the latter a probabilistic formula for the heavy system to
interact is usually employed. This formula can also be derived in a field
theoretical approach.
All these
features are discussed in \cite{Braun}. 
In this way the situation is different in open charm as well as in
Drell-Yan (DY) 
production, where there is no absorption of the produced heavy system (at
least at moderate energy) but the effect of nuclear modifications on structure
functions is almost the same as in the case of charmonium \cite{yuli}. 
A combined study of the different
reactions should help to distinguish the contributions of each effect to
the suppression.

In this paper we are only interested in studying the effect of the
modification of the nucleon structure functions inside nuclei (EMC effect). 
As we will see, this modification has a
big influence on the absorption of charmonium
at large $x_F$, a problem that remains unsolved \cite{kopel}. 
Although this effect is negligible in the integrated cross section at present
energies, it will be important at higher ones. 
A better understanding of
the different suppression mechanisms in pA and AB collisions could
shed some light on what is happening in PbPb, and, for that, the 
$x_F$ dependence of
absorption will be very helpful. Concerning suppression, large $x_F$ 
is related with
higher energies in two senses: first, from the point of view of suppression by 
EMC effect, as the momentum fraction of the target parton
that takes part in the collision is
$x_2\simeq M^2/sx_F$, larger $x_F$ as well as $s$ imply smaller $x_2$ 
(the region
of shadowing corrections). Second, from the point of view of rescattering of
produced pre-resonances 
by nuclear matter (not studied here), again larger $x_F$ or 
$s$
means faster produced particles, that could lead to a bigger absorption cross
section \cite{abscs,abscs2}. This is discussed in detail in reference
\cite{Braun}. 

Further uncertainties come from the treatment of color in absorption of
pre-resonant state by 
nuclear matter. It is usually assumed that a pre-resonant color octet
state is absorbed by rescattering with other nucleons inside nuclei. But, as it
is well known by FA, $J/\Psi$ is not fully formed in color octet state and, 
as we will see, the fraction of the different color contributions to the
production cross section 
is affected by the EMC effect, being the color singlet states less
shadowed than the color octet ones. Then, although the correction by EMC
effect itself is not very big, it will have some effect in the amount 
of absorption by
nuclear rescattering if different absorption cross section are used to color
octet and color singlet states as done, for example, in reference \cite{GV}.

\section{QUARKONIUM PRODUCTION CROSS SECTION}

The production cross section for a quarkonium state $H$ according to the process

$$
P+T\rightarrow H+X
$$

\noindent
can be described in framework of parton model by the integral
\begin{equation}
\sigma_H=\int dx_1dx_2\sum_{i,j}f_i^P(x_1)f_j^T(x_2)\hat\sigma(i,j\rightarrow
H) ,
\label{pmod}
\end{equation}

\noindent
where $f_i^{P(T)}(x_1)$ is the i-parton distribution function in the projectil
(target).

The FA based on NRQCD gives
the partonic elementary cross section,

\begin{equation}
\hat\sigma(i,j\rightarrow H)=\sum_n C^{ij}_{\bar QQ[n]}<{\cal O}^H_n>,
\end{equation}
in terms of short distance parts $C^{ij}_{\bar QQ[n]}$ describing the partonic
process to obtain a $\bar QQ$ pair in a state $[n]$ and long distance
coefficients $<{\cal O}^H_n>$ which describe the hadronization of this state 
into a
physical particle $H$. The short distance parts can be calculated in series of
$\alpha_s(2m_Q)$. However, the long distance ones are non-perturbative 
quantities,
so they are not exactly calculable at present. They can be taken as parameters 
to be
fixed by experimental data or lattice calculations. The matrix elements for
the production of charmonium and bottonium states were computed in 
\cite{BR}-\cite{GS}.

One of the quantum numbers which describe the state $[n]$ is the color of the
$c\bar c$ pair. Then, in this approach, contributions coming from different 
color states of the pre-resonant $c\bar c$ pair are separately taken into 
account.

An alternative model for quarkonia production is the 
CEM \cite{CEV,halzen}. In this model it is assumed that the
different quarkonia states come from the $\bar QQ$ pairs created below the
threshold for open charm production (i.e. $\hat s\equiv x_1 x_2 s=4m_D^2$). 
In this
case the cross section for the production of a given particle $H$ is assumed to
be a constant fraction $F^p_H$ of the hidden cross section 
$\tilde\sigma_{\bar QQ}$
defined as: \cite{CEV}
\begin{equation}
\tilde\sigma_{\bar QQ}(s)=\int_{4m_c^2}^{4m_D^2} d\hat s\int dx_1 dx_2
f_i^P(x_1)f_j^T(x_2)\hat\sigma(\hat s)\delta(\hat s-x_1x_2 s) ,
\label{hidden}
\end{equation}

\noindent
so that
\begin{equation}
\sigma_{pN\to H}(s)=F^p_H \tilde\sigma_{\bar QQ}(s),
\end{equation}

\noindent
being $F^p_H$ the parameter for each particle $H$. This parameter is 
free and independent of the process. $\hat\sigma(\hat s)$ in equation 
(\ref{hidden}) is the partonic cross section to produce a pair $c\bar c$ with
and invariant mass $\hat s$ calculated in pQCD \cite{NDE}. 

For $J/\Psi$ production,
contributions coming from the different decays must be taken into account.
At fixed target energies the
main contributions are the decays of the various $\chi_{cJ}$ states and that of
the $\Psi'$. Other contributions, like the decay of $B$ states are negligible
\cite{BR}. Thus,
\begin{equation}
\sigma_{J/\Psi}=\sigma_{J/\Psi}^{dir}+B(\Psi'\to J/\Psi)\sigma_{\Psi'}+
\sum_{J=0,1,2} B(\chi_{cJ}\to J/\Psi)\sigma_{\chi_{cJ}},
\end{equation}

\noindent
being $\sigma_{J/\Psi}^{dir}$ the direct $J/\Psi$ production, i.e. not coming
from decays, and $B(H\to J/\Psi)$ the branching ratios for particle $H$ to
decay into a $J/\Psi$. Experimentally it is found that 
the direct $J/\Psi$ contribution is about 60\% of the
total $J/\Psi$ cross section, $\Psi'$ decay is less than 10\% and all 
$\chi_{cJ}$
contribute with more than 30\%.

Decays into $\Psi'$ are not important, and then $\Psi'$ production is 
dominated by direct production.

In CEM it is not necessary to make this distinction, because the decays are
implicitly
taken into account in the parameters $F^p_H$. The FA on the contrary offers
the possibility to study these different contributions in a separate way,
calculations in \cite{BR}-\cite{GS} give an accurate description 
of these ratios. 

Another important point is the contribution of the different color states to the
production of these particles, i.e. the color content of the pre-resonant state.
In fact, the FA gives that 
direct $J/\Psi$ production is almost completely produced in color octet state,
$\Psi'$ is also 
predominantly (about 90\% or more) produced in color octet, and
the main contribution of the $\chi_{cJ}$ states to $J/\Psi$ are given by color
singlet matrix elements. Then, a separate study of the EMC effect in different
particles and color states is possible in this approach. 
As we will see, the color octet and color singlet
contributions to the production of charmonium have different suppression by EMC
effect.  

To introduce nuclear corrections by the EMC effect, we use the
results of \cite{ESK}. In this reference a parametrization of experimental data
for the ratio 
\begin{equation}
R_i^A(x,Q^2)={f_i^A(x,Q^2)\over A f_i^N(x,Q^2)}
\label{ratio}
\end{equation}

\noindent
is evolved by means of DGLAP equations. For this parametrization only DIS and
DY experimental data has been taken into account. 
In this formula, $A$ is the atomic
number, $f_i^A(x,Q^2)$ is the i-parton distribution function for nucleons 
inside nucleus
$A$ and $f_i^N(x,Q^2)$ is the corresponding one in a free nucleon. Thus, the
structure functions of equation (\ref{pmod}) must be substituted by the
corresponding ones for nucleons inside nuclei: $A f_i^N(x,Q^2) R_i^A(x,Q^2)$.

These ratios have a shadowing region for $x\lesssim 0.1$, in which the parton
distribution function inside nucleus is depleted ($R_i^A < 1$)
in comparison with that of
free nucleon, an antishadowing region for $0.1\lesssim x\lesssim 0.3$ 
in which this ratio is
bigger than 1, a second shadowing region for 
large $x$ and a second antishadowing region (usually atributed to Fermi Motion
of nucleons)
for $x\gtrsim 0.85$.

\section{NUMERICAL CALCULATIONS}

The cross sections in the FA have been calculated using CTEQ-LO \cite{CTEQLO}
parton distribution functions, $m_c=1.5$ GeV and $\mu=2m_Q$, i.e. the same
parameters as in \cite{BR} 
in order to use the same values for the 
matrix elements $<{\cal O}^H_n>$
fitted by them. In the case of CEM we use GRV-HO \cite{GRV} parton distribution
functions and the same values for the c quark mass and renormalization scale.

The results are presented in the usual parametrization\footnote{It is worth
noting that this common parameter $\alpha$ is in fact dependent on A and then,
although useful, it must be taken with caution.}
\begin{equation}
\sigma^{pA(AA)}=A^\alpha \sigma^{pp}.
\label{paramet}
\end{equation}

It is clear that CEM will give the same results for all the different charmonium
states, because the factors $F^p_H$ are canceled when the ratio is done. It is
not so in the case of FA, where every particle has different contributions
coming from the different 
matrix elements and in the case of $J/\Psi$ also from the decays of the other
charmonium states. 

In Fig.1 it is shown the $\alpha$ parameters for the production of
total $J/\Psi$, direct
$J/\Psi$, $\Psi'$ and the contribution of the various
$\chi_{CJ}$ states to the $J/\Psi$ in proton-gold collisions. 
Different color contributions are presented
in separate curves, being each curve the result of (\ref{paramet}) in which the ratio
is done taking the same contribution of color (i.e.
$\sigma^{pA}_{singlet}/\sigma^{pp}_{singlet}$,
$\sigma^{pA}_{octet}/\sigma^{pp}_{octet}$ and
$\sigma^{pA}_{total}/\sigma^{pp}_{total}$). 
We can see
here the general features of the nuclear effects in charmonium production. First
of all, the results for the CEM and for the FA are not so different, and 
difference  is much smaller than the absorption produced by other mechanisms, 
like nuclear
absorption of pre-resonant states by nuclear matter or co-movers.
Second, shadowing 
corrections are more important at higher energies, as it is expected, because
predominant contribution to the cross sections comes from 
partons in the target with
$x_2\simeq M^2/s$,
so as energy raises, we are going to lower and lower values of $x$, 
the region of larger shadowing. Finally, color singlet states are less shadowed
than color octet ones (color singlet curves
are always above color octet ones). Moreover, more than 90 \% of 
direct $J/\Psi$ is produced in
color octet, while, when different charmonium decays into
this particle are taken into account, this ratio is less than 70\%. This gives
direct $J/\Psi$ production more suppressed than total one.
In Fig.2 we present our results for gold on
gold collisions. The same general trends as in proton-gold are present in this
case. We can observe that effects now are comparatively less important than in
the proton-gold case.

We have also calculated the $x_F$ dependence of nuclear effects for these
processes. 
Figs. 3 and 4 are the $\alpha$'s for 
proton-gold and gold-gold collisions for
$\sqrt{s}=39 GeV$. We find that shadowing corrections are much more important
as $x_F$ is getting larger, as expected (see above). 
This fact is found experimentally. 
Direct comparison of our results with experimental data is not possible because
rescattering of produced pre-resonances is the main contribution to absorption.
However, we are allowed to compare how much of the experimentally found bigger
absorption at large $x_F$ is due to shadowing in the structure functions. To do
this, we consider rescattering corrections to be independent on $x_F$. If both
effects are taking into account in separate exponents

\begin{equation}
\sigma^{pA}=A^{\alpha(x_F)}A^{\beta-1}\sigma^{pp},
\end{equation}
\noindent
where $\alpha(x_F)$ are our results and $\beta$ is a new absorption parameter
independent on $x_F$ and parametrizing other effects different from EMC,
we can compare our results with experimental
data by normalizing our curves in this way 
to one of the less shadowed experimental points.
This has been done in  
Figure 5 where experimental points are from reference \cite{exp}\footnote{In
this 
reference different targets were used and a parametrization in the form (\ref{paramet})
is given taken into account all the different nuclei. As we have said, this parameter is
in fact A-dependent, then for comparison with our calculations only data for W target is
used (we supose that nuclear corrections to PDF in Au and in W are almost the same).}. This normalization corresponds to $\sigma^{abs}\sim$ 6 mb in the
probabilistic formula for nuclear absorption, which is smaller than
$\sigma^{abs}\sim$ 7mb, used for instance in \cite{Nestor} and \cite{qa}.
A different shape is
found for large enough $x_F$ (not reached for $\Psi'$), 
being the experimental data more absorbed. Then EMC
effect alone can not be responsible for this suppression, the additional
depletion coming from other effects like an increase of 
rescattering absorption
at large $x_F$, intrinsic charm, etc. 
The study of the EMC effect in charmonium production has been 
done previously in references \cite{GSatz,ChL,GHQ}. 
In this
references, scaling in $x_2$ of the ratios $\sigma^{pA}/\sigma^{pp}$ 
is found for some energies. This scaling is ruled out in reference \cite{exp}.
In our approach this scaling violation should be due a different final 
state absorption depending on $x_F$ and $s$.
Reference \cite{GSatz} restrict their study to the $0<x_F<0.55$ region. In this
sense, our results are in agreement with theirs. The main differences between
the present work and the previous ones are the use of charmonium production
cross section taken from hadron-hadron collisions and parametrization of
EMC effect from DIS and DY experimental data.
In particular, in \cite{ChL} the parametrization for the ratio (\ref{ratio})
is much more shadowed than the one used by us for low $x$ and the
$\sigma^{abs}$ used there to reproduce experimental data is quite small
($\sigma^{abs}=$2.6 mb).
%is in
%contradiction with experimental data of DIS at small $x_2$.

\section{CONCLUSIONS}

Effects of the modification of nucleon parton distribution functions inside
nucleus on the production of different charmonium states have been studied.
Energy as well as $x_F$ dependence of this effects have been calculated for
proton-gold and gold-gold collisions. We have found a depletion on the ratio
between the pA(AA) to pp cross sections for higher energies and larger $x_F$.
This effect is, at present energies, 
much less important that the other absorption mechanisms
of charmonium production, like pre-resonant absorption by rescattering
with other nucleons in the nucleus or by co-movers. 
However, these
processes superimpose to the production mechanism, so they are affected by the
corrections due to EMC effect. 

When FA is used, a separate study of the contributions of 
different color states and particle
decays to the production mechanism is possible. 
Our result is that color octet
states are more shadowed than color singlet ones, and that a slight difference
is obtained for $J/\Psi$ production when decays of
other charmonium states are taken into account.

We have also 
found a slightly different absorption for $J/\Psi$ and $\Psi'$ states,
in the case of FA.
This difference is not present in experimental data, being too small to be
seen over the other absorption mechanisms. We can say then that nuclear
corrections for $J/\Psi$ and $\Psi'$ production due to EMC effect are almost
the same. In the CEM, as we have already said, 
they are exactly the same. We also obtain
that CEM and FA give almost the same results in all our calculations.

Finally, we obtain 
a different shape for experimental and theoretical curves in $x_F$,
being the suppression in experimental data bigger than 
that in our 
calculations for large $x_F$. 
Although our curves almost fit the experimental points (within errors) for the
$x_F$ dependence of nuclear suppression, data seems to prefer a bigger
absorption for large $x_F$. So we can conclude that modification of the nuclear
parton distribution functions can account for the $x_F$ dependence of 
$J/\Psi$ suppression for moderate $x_F$ and that other  
mechanisms apart from EMC effect is needed at large $x_F$ in order to reproduce
experimental data.

{\bf Acknowledgements}

We are grateful to N. Armesto and A. Capella 
for useful discussions and to
K. J. Eskola for sending us their numerical results. We thank the Direcci\'on
General de Pol\'\i tica Cient\'\i fica and CICYT for financial support under
contract AEN96-1673. C.A.S. also thanks Xunta de Galicia for a grant.

\pagebreak

\pagebreak
\vskip 2cm

\begin{center}
{\bf Figure captions}\\
\end{center}
\vskip 1cm
\noindent
{\bf Fig.1}. Center of mass energy dependence of $\alpha$ for p-Au collisions. 
Different lines are:
FA total contribution (solid), singlet contribution (dotted), octet
contribution (dashed) and CEM (dotted-dashed)

\vskip 1cm
\noindent
{\bf Fig.2}. Same as Fig.1 but for Au-Au collisions.

\vskip 1cm
\noindent
{\bf Fig.3} $x_F$ dependence of $\alpha$ for $\sqrt{s}=39$ GeV 
p-Au collisions with the same
conventions for lines as in Fig.1.

\vskip 1cm
\noindent
{\bf Fig.4}. Same as Fig.3 but for Au-Au collisions.

\vskip 1cm
\noindent
{\bf Fig.5}. Comparison of $x_F$ dependence of nuclear corrections by EMC
effect with experimental data of 800 GeV/c protons incident on tungsten target from
reference \cite{exp}. Theoretical calculations have been normalized to
one of the less shadowed experimental points (see text). 
Lines follow the same convention as Fig.1.

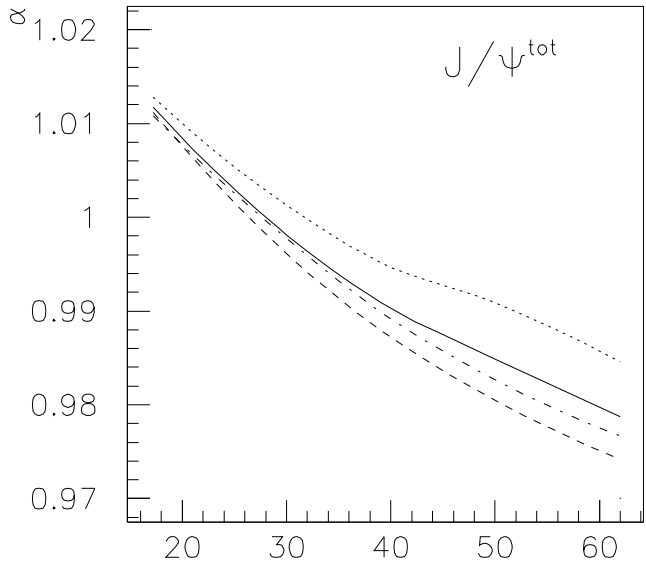
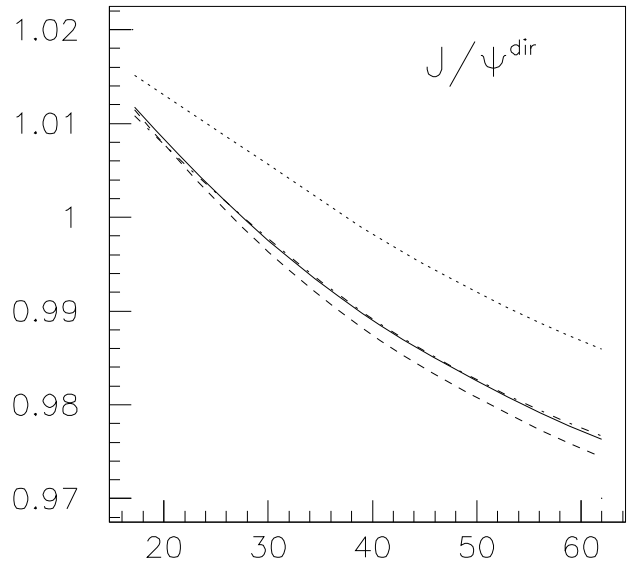
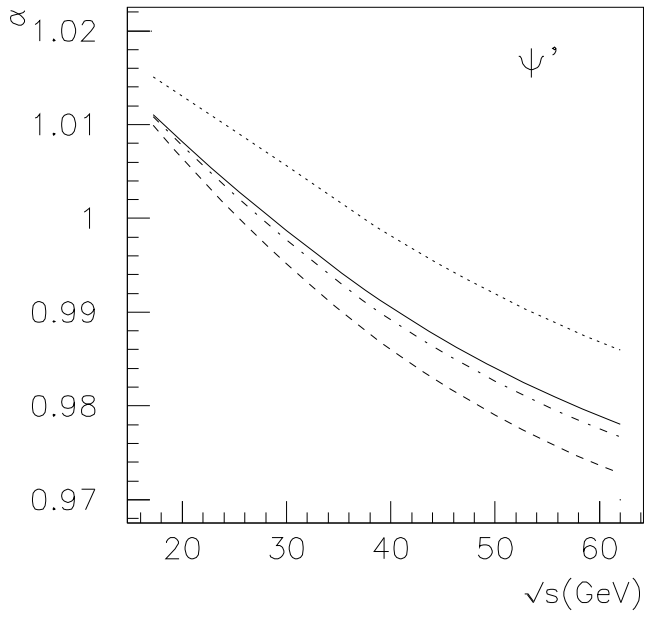
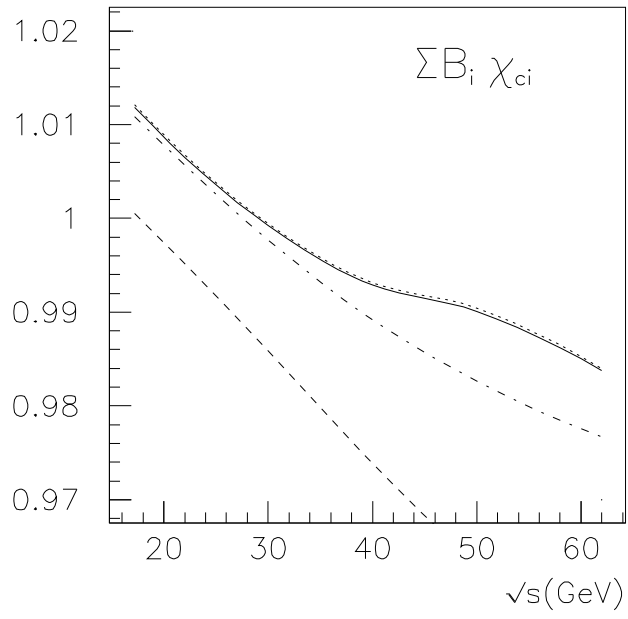
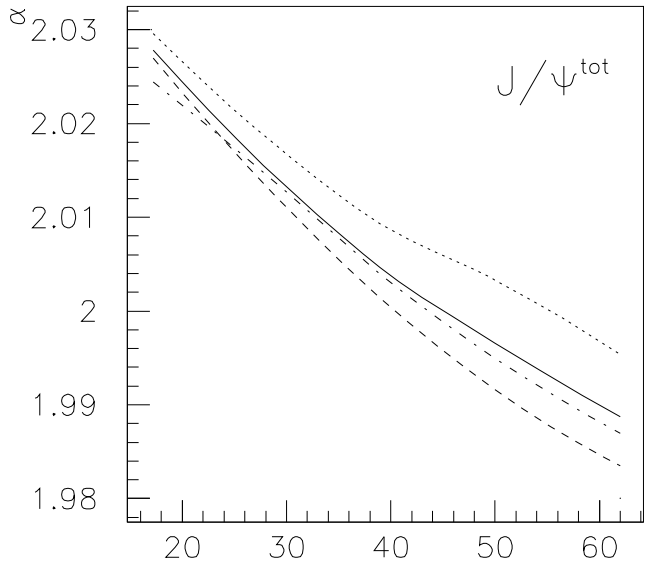
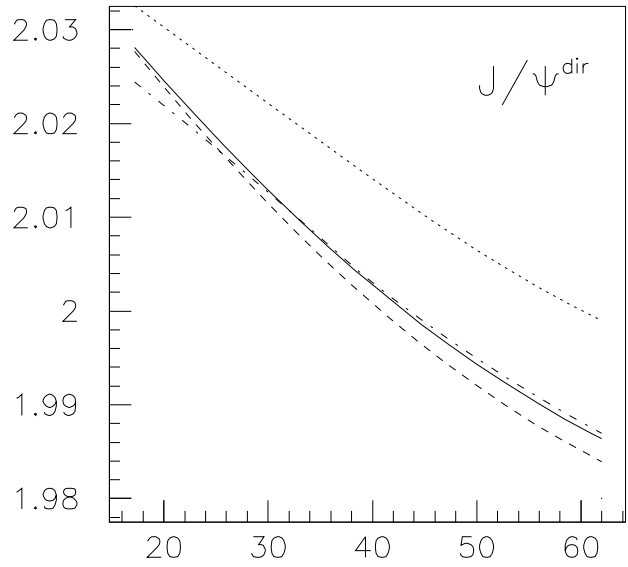
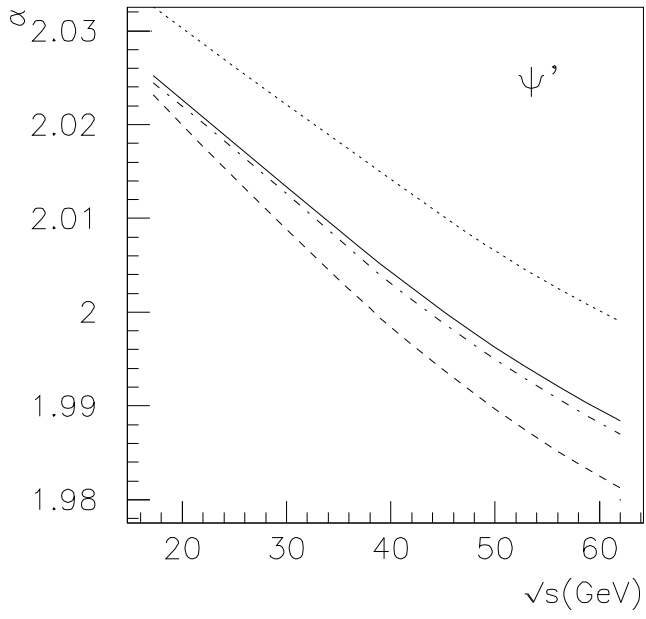
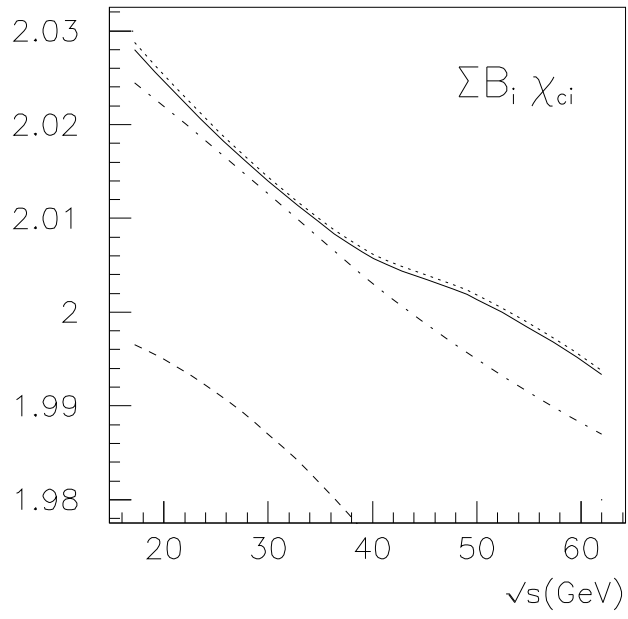
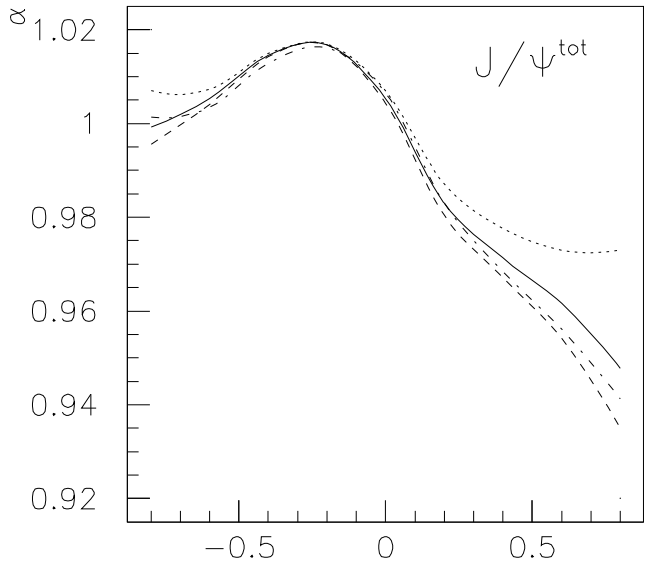
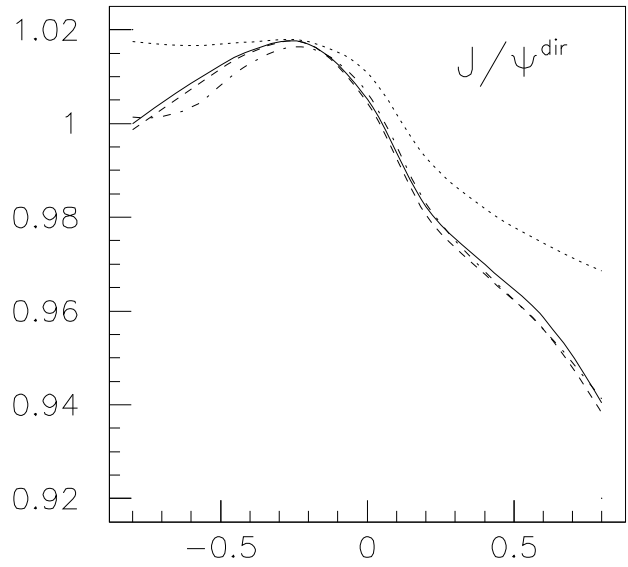
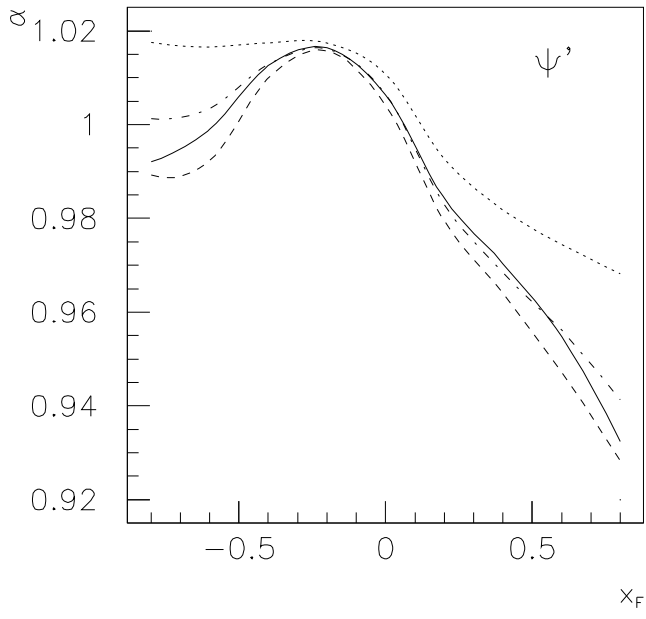
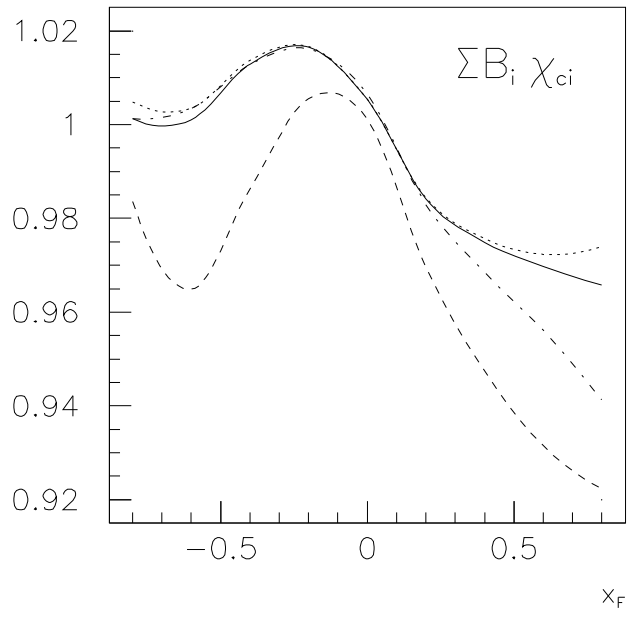
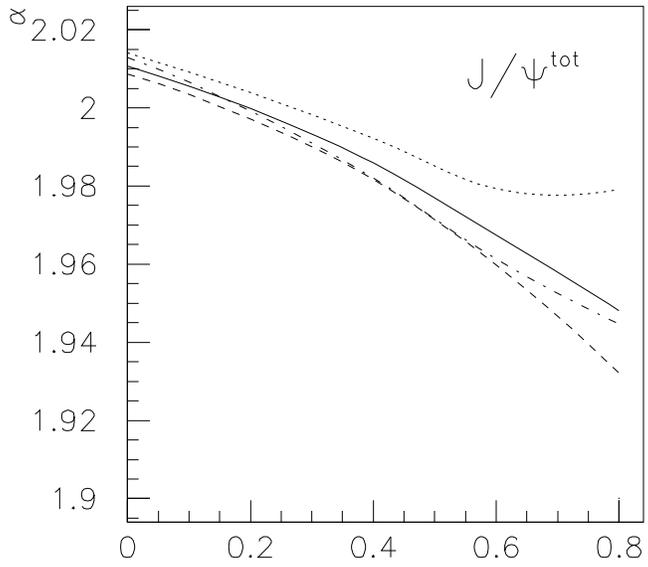
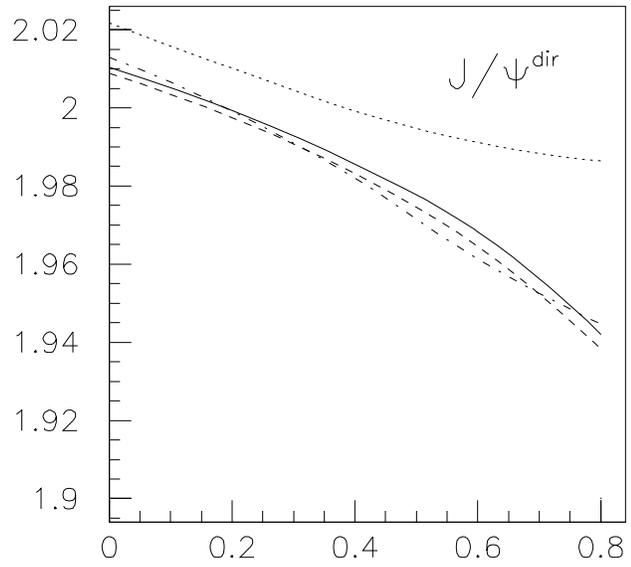
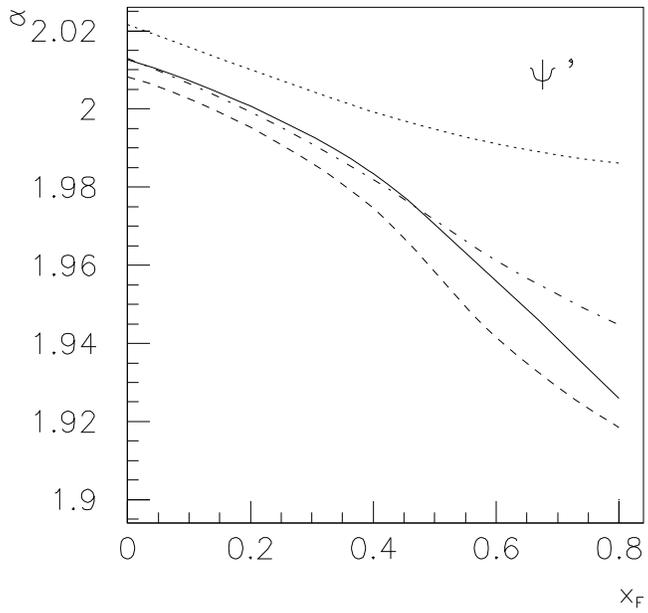
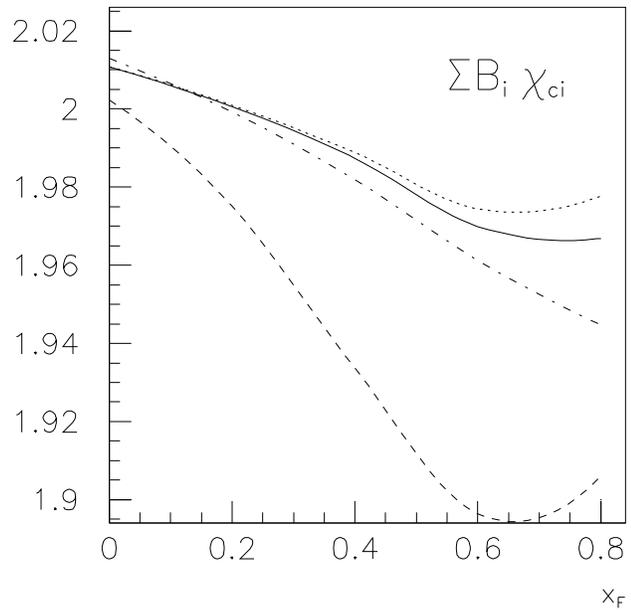
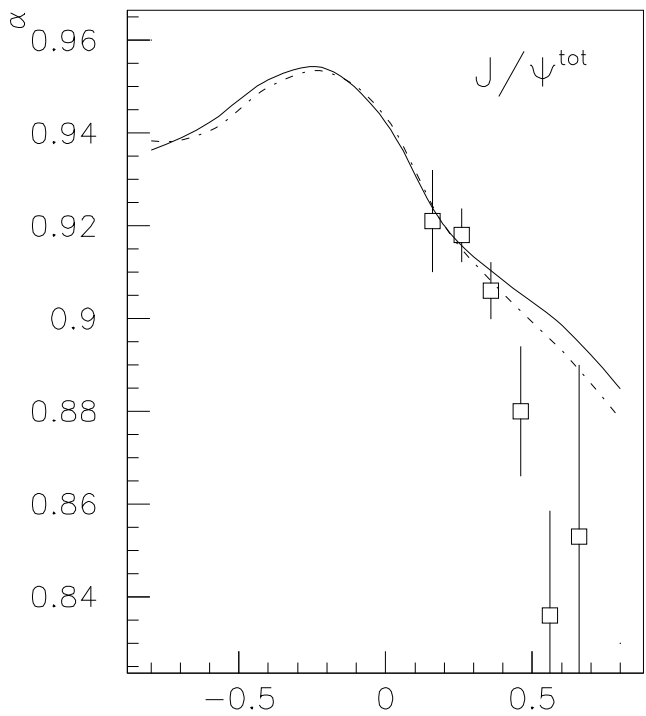
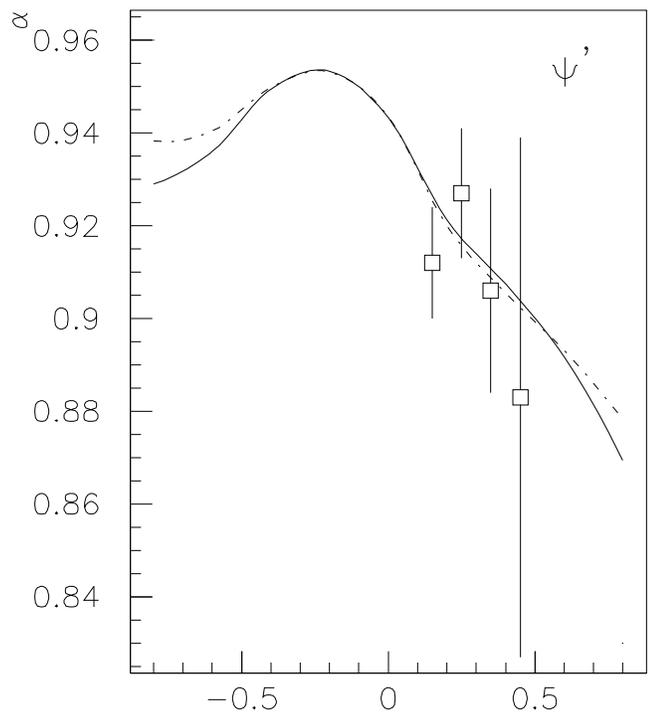
\end{document}